\documentstyle[12pt,epsfig]{article}
\renewenvironment{thebibliography}[1]
	{\begin{list}{\arabic{enumi}.}
	{\usecounter{enumi}\setlength{\parsep}{0pt}
	 \setlength{\itemsep}{0pt} 
         \settowidth
	{\labelwidth}{#1.}\sloppy}}{\end{list}}

\newcommand{\scaption}[1]{\caption{\protect{\footnotesize  #1}}}

\newcommand{\av}[1]{\mbox{$ \langle #1 \rangle $}}

\newcommand{\xb}{\mbox{$x~$}}  

\newcommand{\Qsq}{\mbox{$Q^2~$}}
\newcommand{\et}{\mbox{$E_T~$}}
\newcommand{\kt}{\mbox{$k_T~$}}
\newcommand{\pt}{\mbox{$p_T~$}}

\newcommand{\ethad}{\mbox{$E_T^{\rm had}~$}}
\newcommand{\etpar}{\mbox{$E_T^{\rm par}~$}}
\newcommand{\ptmax}{\mbox{$p_T^{\rm max}~$}}

\newcommand{\as}{\mbox{$\alpha_s~$}}

\newcommand{\GeV}{\mbox{\rm ~GeV~}}

\newcommand{\GeVsq}{\mbox{${\rm ~GeV}^2~$}}

\begin{document}

\begin{titlepage}

%
\noindent
{\tt MPI-PhE/96-10} \\
{\tt hep-ph/9606246} \\
{\tt June 1996}                  \\

\begin{center}

\vspace*{2cm}


{\bf  A NEW METHOD TO PROBE THE LOW 
$x$ PARTON DYNAMICS  \\ AT HERA} \\
%
\vspace*{2.cm}
{\bf M. Kuhlen} \\ 
\vspace*{1.cm}
Max-Planck-Institut f\"ur Physik \\
Werner-Heisenberg-Institut \\
F\"ohringer Ring 6 \\
D-80805 M\"unchen  \\
Germany \\
E-mail: kuhlen@.desy.de
\bigskip
\bigskip
\\

\vspace*{2cm}

\end{center}

\begin{abstract}

\noindent 
Hadron transverse momentum spectra are
proposed as a means to probe 
the underlying partonic dynamics
in deep inelastic scattering.
The BFKL evolution equation,
postulated for small Bjorken-$x$,
leads to an enhanced
parton emission over the conventional DGLAP ansatz, 
and can thus be tested.

\vspace{1cm}

\end{abstract}
\end{titlepage}

\newpage

\section{Introduction}

\begin{figure}[htp]
   \centering
   \vspace{-0.2cm}
   \epsfig{file=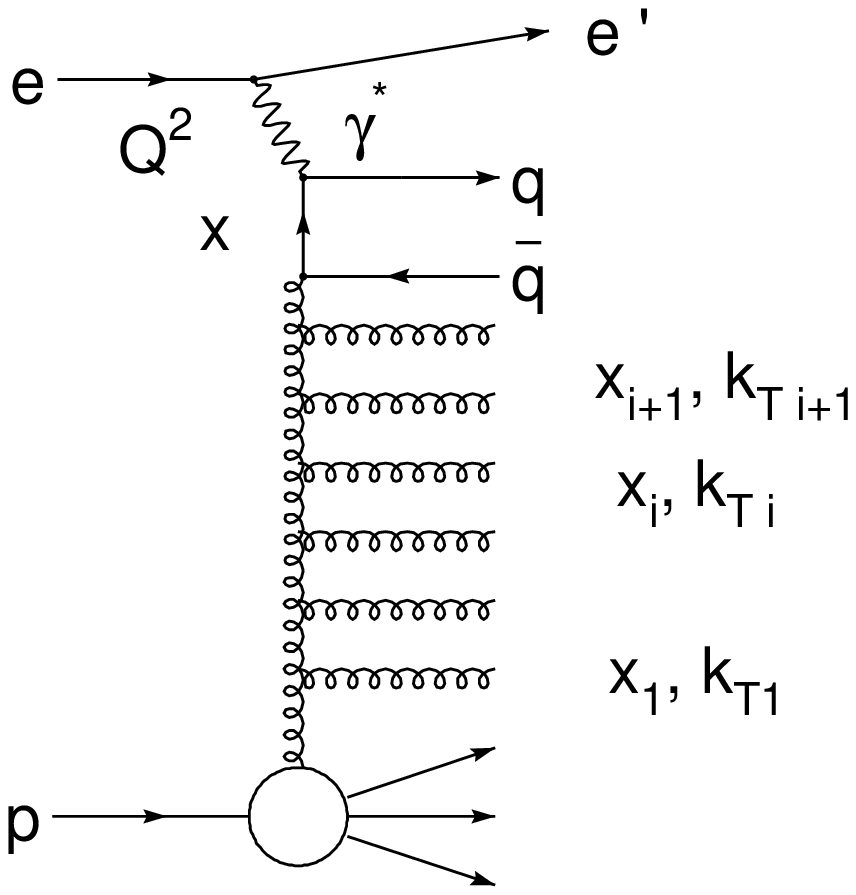,width=5cm}
   \scaption{ 
      Parton evolution in the 
      ladder approximation. The longitudinal
      fractional momenta $x_i$ and transverse momenta $k_{Ti}$
      of subsequently emitted partons are indicated.   }
   \label{cascade} 
\end{figure}

The successful description 
of the nucleon structure function data 
by perturbative QCD, cast into the
DGLAP (Dokshitzer-Gribov-Lipatov-Altarelli-Parisi) 
parton evolution 
equations \cite{dglap} constitutes
one of the major successes of QCD. 
At small enough Bjorken-\xb however,
these equations are expected to break down. 
An alternative ansatz
for the small \xb regime is the BFKL 
(Balitsky-Fadin-Kuraev-Lipatov) equation \cite{bfkl}.
At lowest order the BFKL and DGLAP equations resum the leading 
logarithmic 
$(\as \ln 1/x)^n$ or $(\as \ln (\Qsq/ Q_0^2))^n$ contributions
respectively, with \Qsq 
being the virtuality of the exchanged photon.
The leading log
DGLAP ansatz corresponds to a strong ordering
($Q_0^2 \ll \kt_1^2 \ll ... \kt_i^2 \ll ... \Qsq$)
of the transverse momenta \kt (w.r.t. the proton beam)
in the parton cascade (Fig.~\ref{cascade})
while in the BFKL ansatz they rather follow
a kind of random walk
($\kt_i^2 \approx \kt_{i+1}^2$)
\cite{ordering}. 
It is an open theoretical question 
to what extent BFKL type contributions
play a r\^{o}le in the small \xb ($\approx 10^{-4}$)
regime now accessible in deep inelastic scattering (DIS) at HERA.
%
%
%
Measurements
on the hadronic final state emerging from the cascade
could be sensitive to the new type of evolution.
For example, without the restriction of strong \kt ordering,
more transverse energy \et is expected  
from BFKL than from DGLAP parton radiation
in a 
region between the current region and the proton remnant
\cite{durham}. 
Though the HERA \et flow data \cite{h1flow} can 
be interpreted consistently with the BFKL mechanism,
it 
was not possible to disentangle 
the perturbative parton radiation from
non-perturbative hadronization effects
\cite{kuhlen,sci}.  

In this paper it will be demonstrated that 
single particle transverse momentum (\pt) spectra 
represent a more direct measure of the partonic activity
than the \et flow measurements. 
Observables are then constructed which allow to discriminate
the \kt ordered from the unordered parton shower scenario. 

Predictions for the cases of the 
ordered resp. unordered cascades are extracted
from Monte Carlo models, which incorporate the QCD evolution in
different approximations and utilize phenomenological models
for the non-perturbative hadronization phase.
The MEPS model (Matrix Element plus Parton Shower) \cite{lepto}, 
incorporates the QCD matrix elements up to first order, with additional
soft emissions generated by adding leading log parton showers.
In the colour dipole model (CDM) \cite{dipole,ariadne}
radiation stems from
colour dipoles formed by 
the colour charges. 
Both programs
use the Lund string model \cite{string} for hadronization.
The Herwig model \cite{herwig}
is also based upon leading log parton showers, with
additional matrix element corrections \cite{seymour}.
It uses a 
cluster fragmentation scheme \cite{cluster}.
The CDM description 
of gluon emission is similar to that of the BFKL evolution, 
because the gluons emitted by the dipoles
do not obey strong \kt ordering~\cite{bfklcdm}. 
In MEPS and Herwig 
the partons
are strongly ordered in \kt, because they are based upon
leading log DGLAP parton showers.
The latest versions of the models
(Lepto 6.4 for MEPS,
Ariadne 4.08 for CDM and Herwig 5.8) are used with
the parton density parametrization MRSH 
\cite{mrsh}. They
provide a satisfactory
overall description of current DIS final state data \cite{carli},
in particular of the \et flows
\footnote{
In MEPS the new concept of 
soft colour interactions \cite{sci} 
had to be introduced to reach the level of \et seen in the data
\cite{h1flow,carli}. Intriguingly,
this mechanism also produces 
rapidity gap events \cite{gap} at a rate comparable to observation
\cite{sci}, roughly 10\%. 
Rapidity gap events are also produced by
the cluster fragmentation in Herwig.
In this paper rapidity gap events are excluded.}.

\section{The Method}

\begin{figure}[t]
   \centering
   \vspace{-0.2cm}
   \epsfig{file=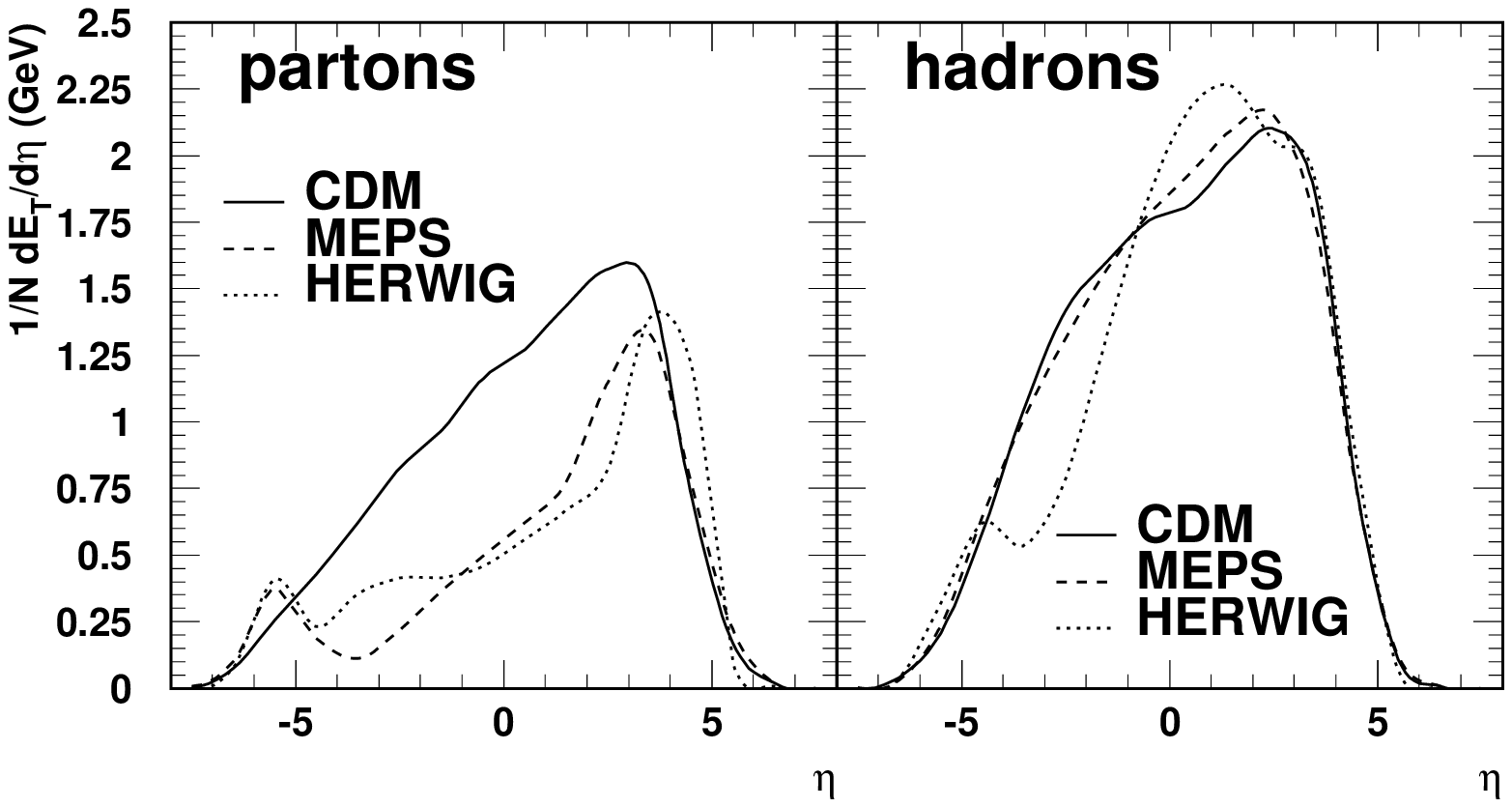,width=12cm,
    bbllx=2pt,bblly=276pt,bburx=518pt,bbury=521pt,clip=}
   \begin{picture}(1,1) \put(-370.,20.){\bf a} \end{picture}
   \epsfig{file=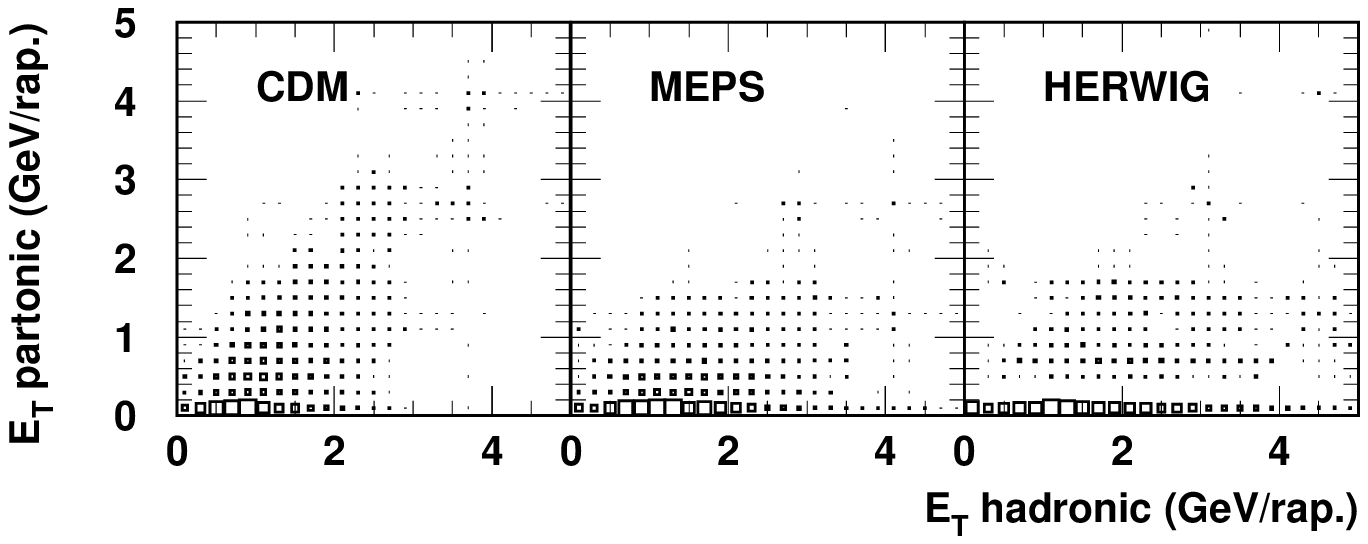,width=12cm,
    bbllx=4pt,bblly=360pt,bburx=514pt,bbury=514pt,clip=}
   \begin{picture}(1,1) \put(-370.,20.){\bf b} \end{picture}
   \scaption{
      {\bf a)} Transverse energy flows vs. pseudorapidity $\eta$
      generated by the models
      CDM, MEPS and Herwig for partons and for hadrons at ``low \xb''.
      The proton direction is to the left.
      {\bf b)} Correlation between partonic and hadronic \et produced
      in the central pseudorapidity bin $0<\eta<2$ for CDM, 
      MEPS and Herwig.}
   \label{etflow} 
\end{figure}

\hyphenation{pre-dominantly}
In this section the sensitivity of single particle \pt spectra
to the parton activity in the ladder is demonstrated.
Generated events are selected from the kinematic
plane of \xb and \Qsq according to the binning chosen by H1 \cite{h1flow}.
Events from two bins, one at ``low \xb'' (\av{\xb}=0.00037)
and one at ``high \xb'' (\av{\xb}=0.0023), with
$\av{\Qsq}\approx 14 \GeVsq$ approximately constant,
are compared.
In Fig.~\ref{etflow}~a the \et flow in the hadronic centre of 
mass system CMS is shown
\footnote{All distributions shown are normalized to the number of
events $N$ which enter the distribution.} 
as a function of pseudorapidity $\eta$ ($\eta=-\ln\tan\theta/2$,
where the angle $\theta$ is measured w.r.t the virtual photon direction) 
for events with small \xb. As expected, the partons produced
from unordered emission (CDM) give more \et in the
central $\eta$ region $\eta \approx 0$
than the ones emitted from the ordered 
cascade (MEPS, Herwig). However, the observable particles emerging 
after hadronization give rise to very similar \et flows,
unresolvable with
current data \cite{h1flow,carli}. While hadronization adds relatively
little \et to the partonic \et for CDM, most of the \et is generated by
hadronization in the cases of MEPS and Herwig. 
To answer the question whether the \et observed in the data 
is generated predominantly
by parton radiation or by hadronization, 
inclusive \pt spectra are considered. 
Hadronization
should produce typical spectra 
which are limited in \pt, while
parton radiation should manifest itself in a hard tail of the \pt 
distribution. 
That tail is due to occasional hard parton radiation,
from which hard particles can emerge.
The production of such hard particles from hadronization
would be 
suppressed.

\begin{figure}[htb]
   \centering
   \vspace{-0.2cm}
   \epsfig{file=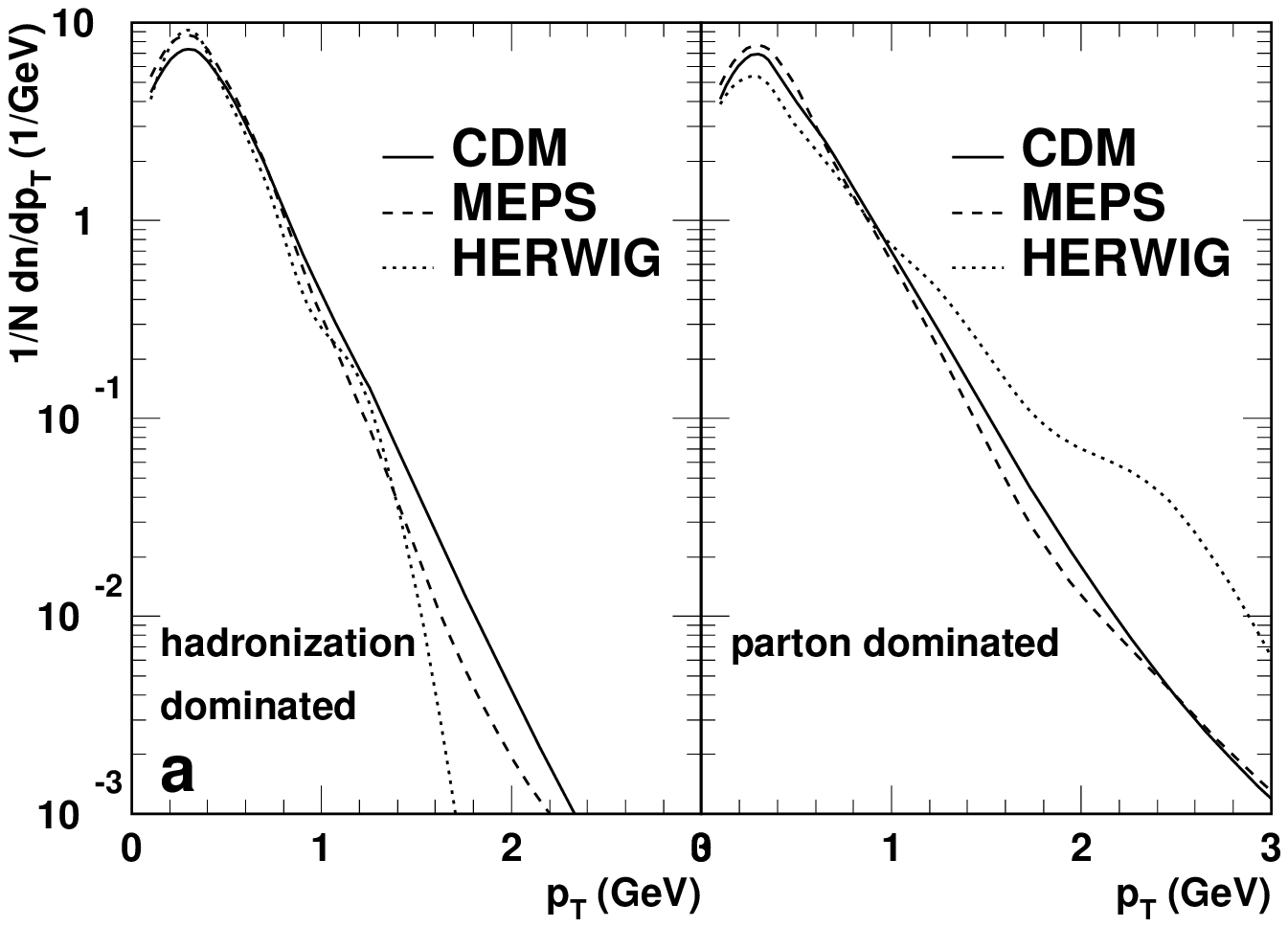,width=10.0cm,
    bbllx=0pt,bblly=0pt,bburx=390pt,bbury=277pt,clip=}
   \epsfig{file=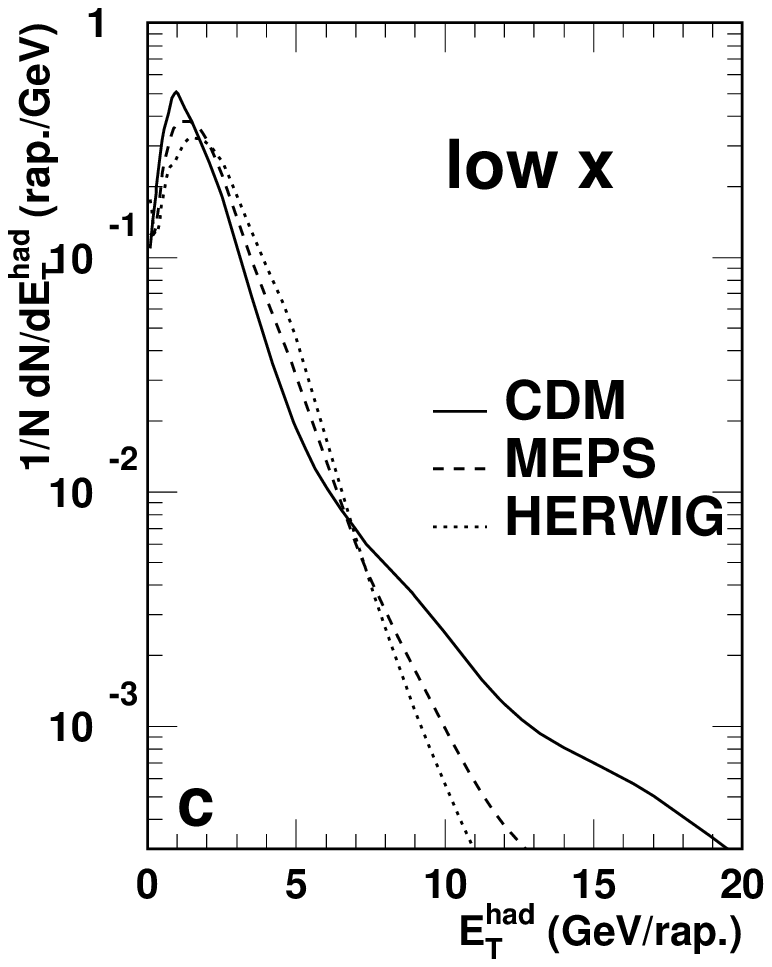,width=5.51cm,
    bbllx=0pt,bblly=0pt,bburx=220pt,bbury=277pt,clip=}
   \epsfig{file=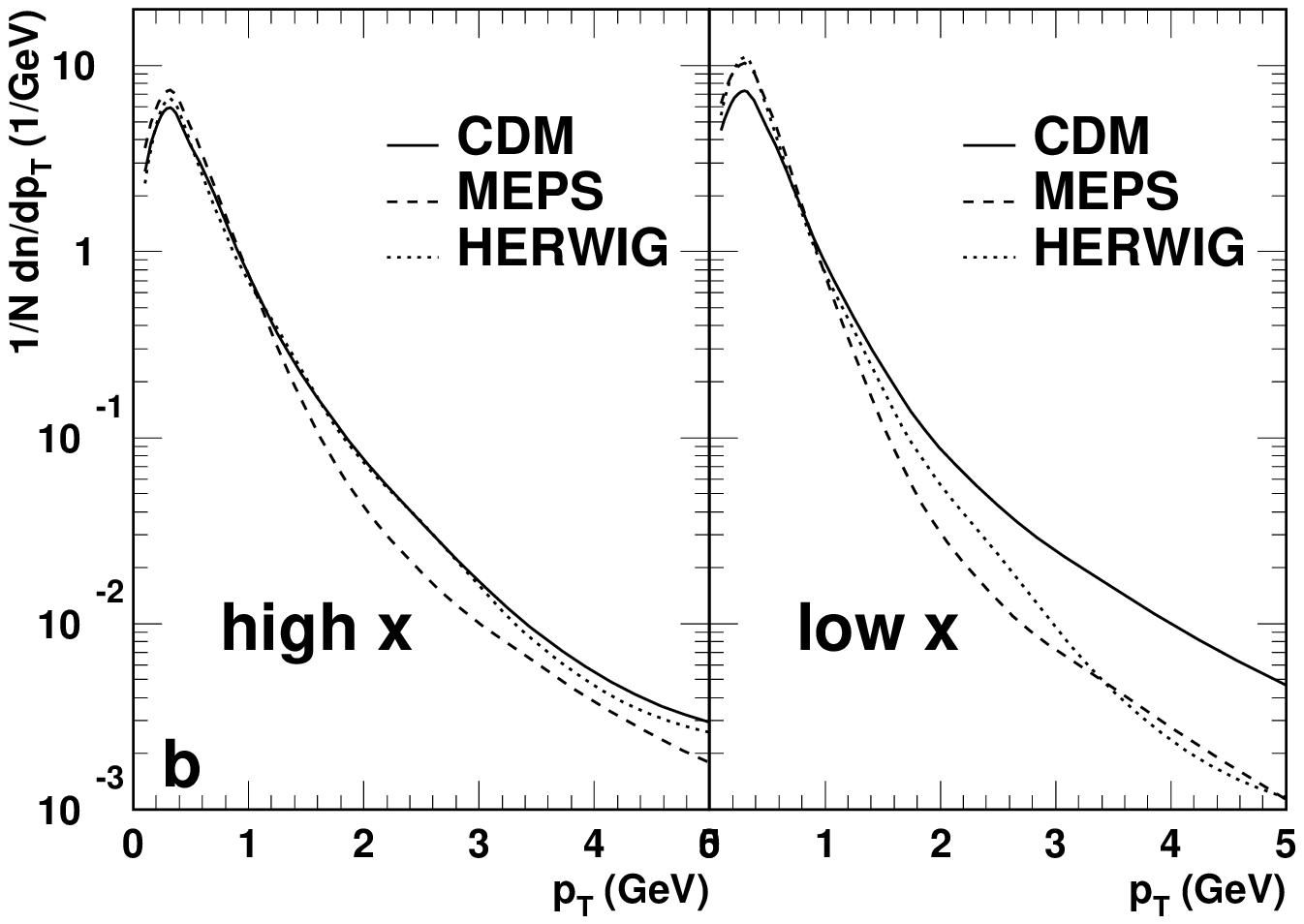,width=10.0cm,
    bbllx=0pt,bblly=0pt,bburx=390pt,bbury=277pt,clip=}
   \epsfig{file=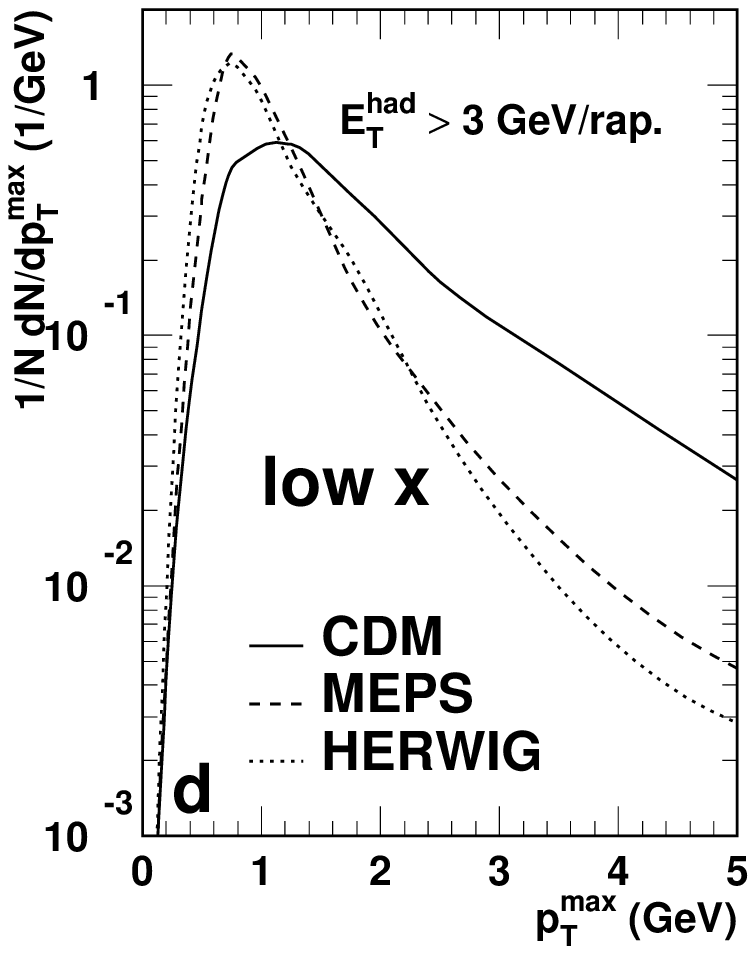,width=5.51cm,
    bbllx=0pt,bblly=0pt,bburx=220pt,bbury=277pt,clip=}
   \scaption{ 
     {\bf a)} Transv. momentum (\pt) spectra of charged particles
      from the central pseudorapidity bin $0<\eta<2$ for events which
      are either hadronization dominated or parton dominated.
      {\bf b)} Ch. particle \pt spectra from  $0<\eta<2$ for 
         events with ``high \xb'' and with ``low \xb''.
%
     {\bf c)} Event distribution in the quantity \ethad, determined from
        $0<\eta<2$, at ``low \xb''.  
      {\bf d)} Distribution of the maximal transv. momentum \ptmax of
      ch. particles from $0<\eta<2$ for
      events with $\ethad>3 \GeV$/unit rap. at ``low \xb''.} 
   \label{ptsens} 
\end{figure}

To test this idea particles from a ``central''
$\eta$ interval $0<\eta<2$ are examined. 
The lower limit is given by the approximate acceptance of the  
HERA detectors, and the upper limit restricts the interval
to the region where the partonic differences in \et are largest,
excluding the ``current'' fragmentation region.
Events are compared which have similar hadronic
\et in that interval 
(\ethad between 1 and 2 GeV/unit rapidity),
but different amounts of partonic \et, \etpar. 
Events 
with $\etpar < 0.2$ GeV/unit rap. are called hadronization dominated,
and events with $\etpar/\ethad > 0.5$ are called parton
dominated. The correlation between \etpar and \ethad is shown in 
Fig.~\ref{etflow}~b. 
For the CDM two classes 
of events can be identified. For one class \ethad is well correlated
with \etpar, for the other \etpar is small, regardless of \ethad. 
For the other models, most events fall into the latter class,
while the correlation between \etpar and \ethad is much less 
pronounced for the rest of the events.
The parton dominated events indeed exhibit a harder \pt spectrum
than the hadronization dominated events (s. Fig.~\ref{ptsens}~a),
regardless of the underlying parton dynamics
or the applied hadronization model. Therefore 
\pt
spectra provide a useful method to study the underlying parton
dynamics in DIS.

\section{Predictions}

In this section
observables are constructed that should allow to distinguish
between the two scenarios of ordered
resp. unordered parton
evolution, or in general be sensitive to the parton radiation
generated in the evolution.
In this study the CDM is taken as a model to represent
the unordered parton cascade, 
and the MEPS and Herwig models
represent the ordered cascade.
In Fig.~\ref{ptsens}~b the inclusive \pt
spectra of charged particles from the ``central'' $\eta$
bin are shown for large and for small \xb.  
At large \xb, all models predict similar \pt 
spectra. 
At small \xb however 
the tail of the distribution 
(\pt larger than $\simeq 1.5\GeV$)
is harder
for the unordered model (CDM) than for the others, as expected
given the larger parton activity.
Less visible due to the logarithmic scale is a difference in
the average charged multiplicity in the central $\eta$ region
between the two scenarios of about 20\%.
In MEPS and Herwig
more soft particles are produced in the hadronization
phase to generate the \et seen in the data. 
%
%

\begin{figure}[t]
   \centering
   \vspace{-0.2cm}
   \epsfig{file=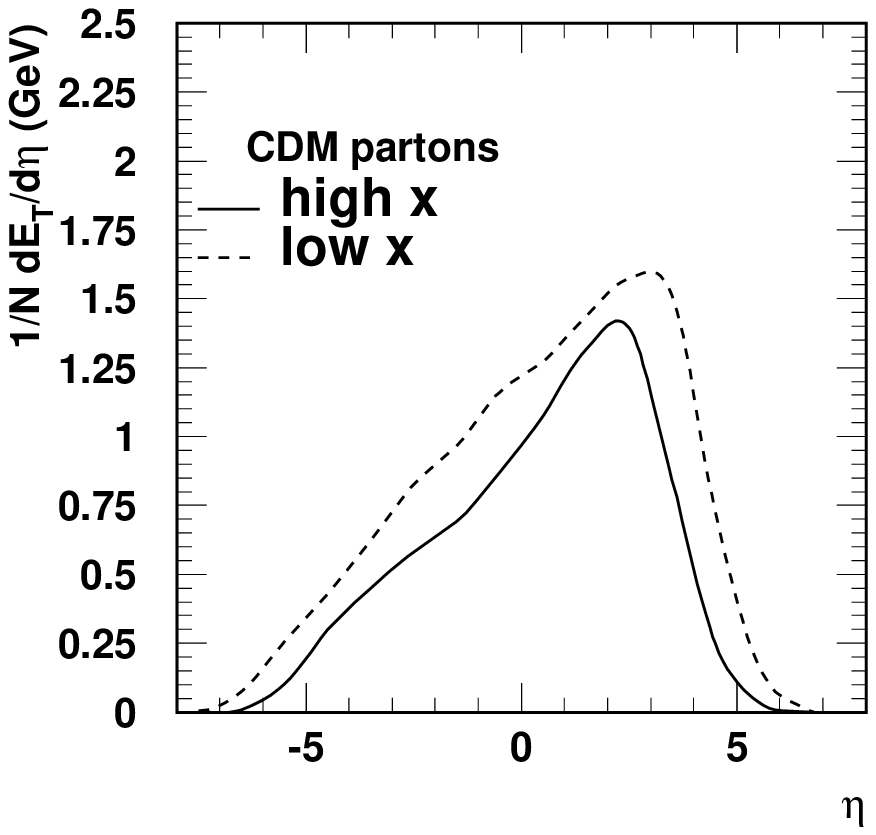,width=5cm,
    bbllx=5pt,bblly=278pt,bburx=258pt,bbury=516pt,clip=}
   \epsfig{file=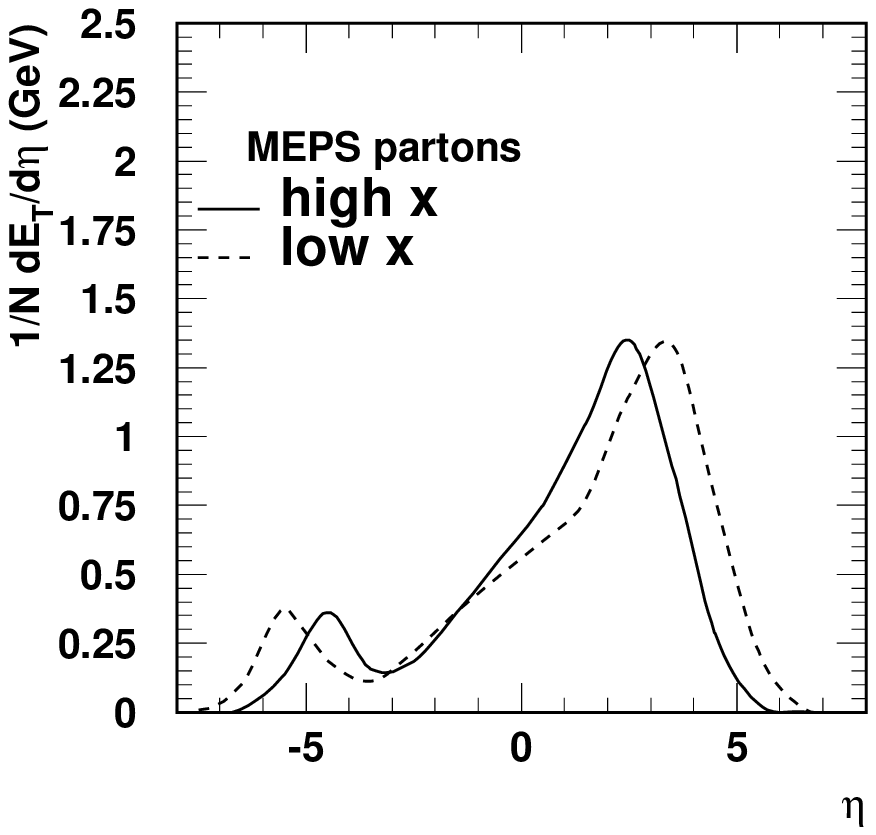,width=5cm,
    bbllx=5pt,bblly=278pt,bburx=258pt,bbury=516pt,clip=}
   \epsfig{file=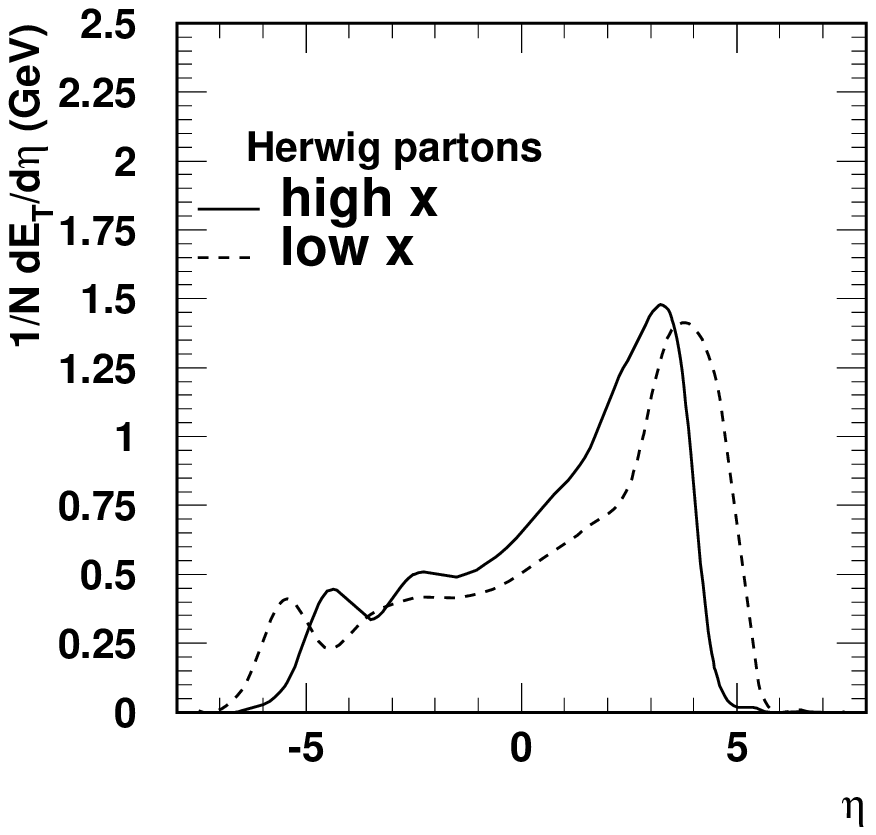,width=5cm,
    bbllx=5pt,bblly=278pt,bburx=258pt,bbury=516pt,clip=}
   \epsfig{file=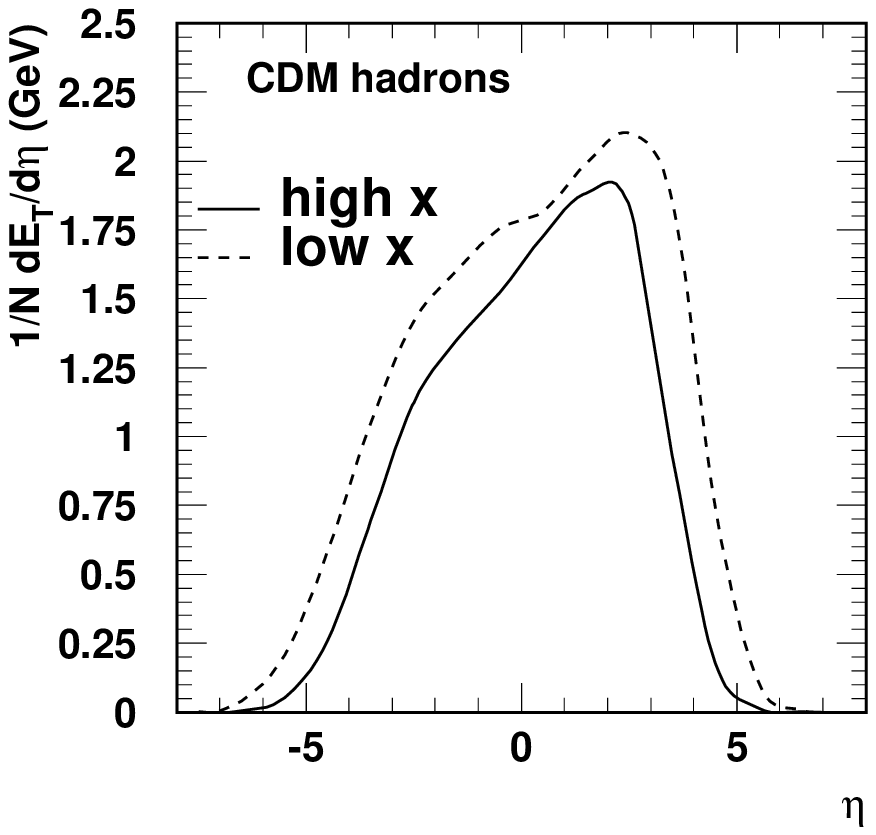,width=5cm,
    bbllx=5pt,bblly=278pt,bburx=258pt,bbury=516pt,clip=}
   \epsfig{file=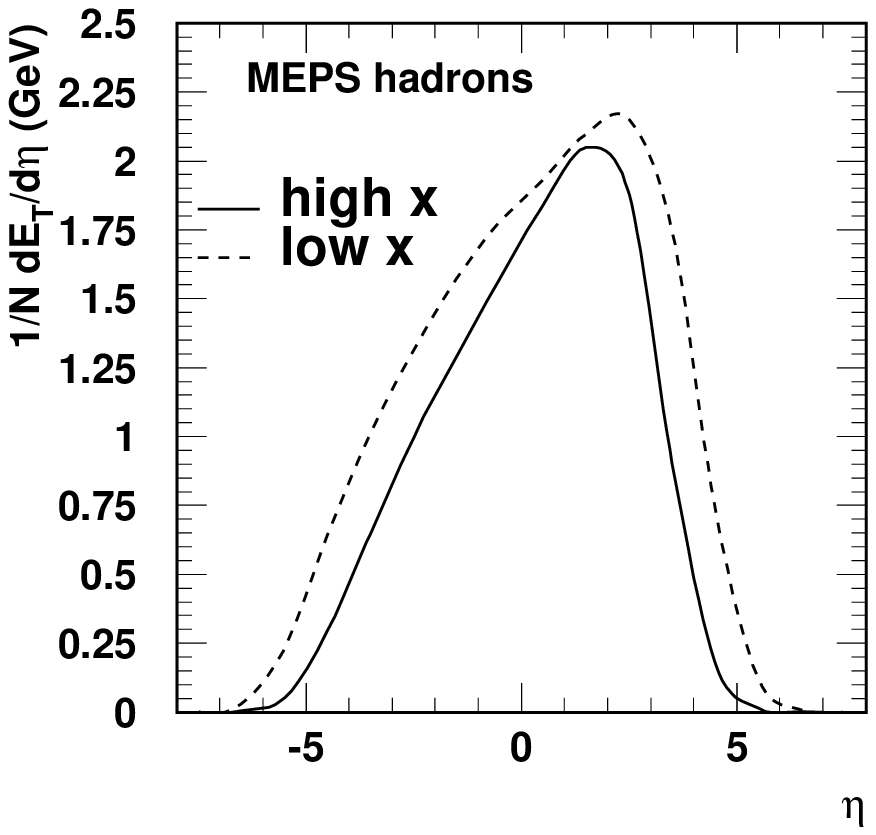,width=5cm,
    bbllx=5pt,bblly=278pt,bburx=258pt,bbury=516pt,clip=}
   \epsfig{file=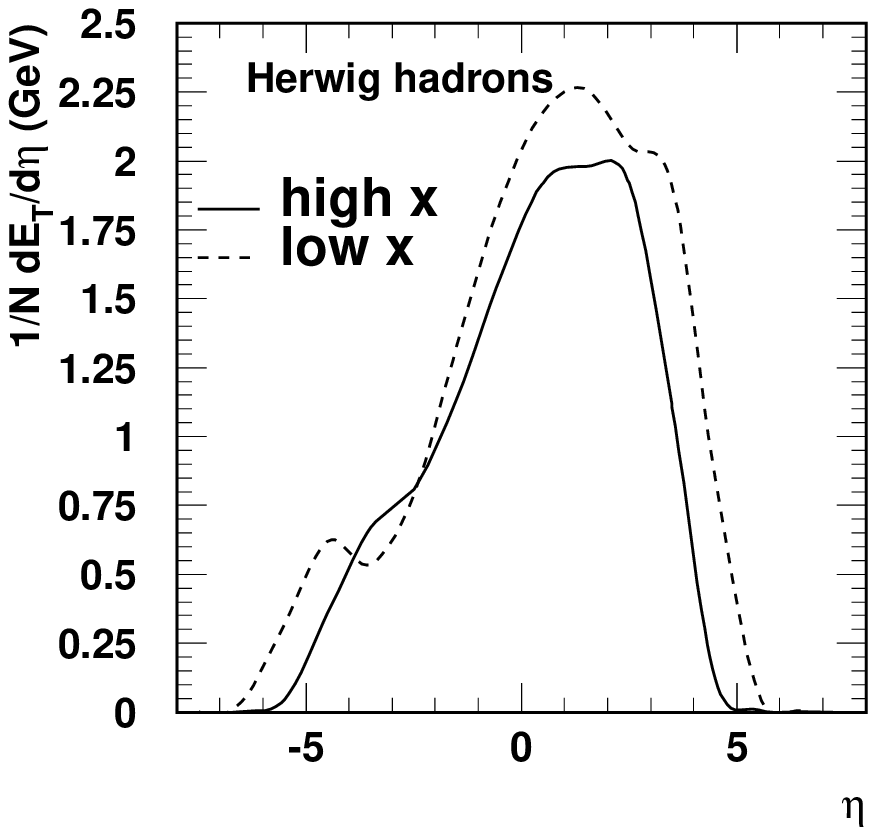,width=5cm,
    bbllx=5pt,bblly=278pt,bburx=258pt,bbury=516pt,clip=}
   \scaption{
      \et flows vs. pseudorapidity for CDM, MEPS and Herwig on
      the parton and on the hadron level. Compared are the \et flows
      for ``high''  and ``low'' \xb.}
   \label{ethilo} 
\end{figure}

One also notices from Fig.~\ref{ptsens}~b
that from the unordered cascade 
a hardening of the spectrum is predicted towards small \xb,
and a softening otherwise. This behaviour
can be traced to the fact that while all models predict an increase
of \ethad towards small \xb, as observed in the data \cite{h1flow},
only the CDM shows that increase also on the parton level
(see Fig.~\ref{ethilo}). 
The other
models predict a decreasing \etpar. This behaviour 
of the models is in accord with
perturbative calculations of the central \et as a function of \xb
\cite{durham}, based upon either the BFKL or DGLAP evolutions. 
As a consequence, the relative amount of
hadronization to \ethad decreases for CDM, but increases for
MEPS and Herwig towards small \xb.
The signal can be enhanced by selecting events in which large
\ethad 
(s. Fig.~\ref{ptsens}~c)
is observed 
(e.g. calorimetrically), 
because in CDM that is correlated
with large \etpar, as opposed to the other models.
For such events a dramatic signal can be obtained by
measuring
the maximal \pt observed in the central $\eta$ region,
s. Fig.~\ref{ptsens} d. 
Enhanced parton radiation would also be signaled in the tail
of the \ethad distribution, s. Fig~\ref{ptsens} c.

%
%

\section{Conclusions}

In order to investigate the dynamical features of parton evolution
in the proton at small \xb, observables based on single particle
\pt spectra have been constructed.
It has been demonstrated for all models investigated
that the hardness
of such spectra is sensitive to parton radiation from the cascade. 
Since for small enough \xb it is expected that 
the DGLAP equations with strong \kt ordering for parton radiation
cease to be valid, and may possibly be substituted by the BFKL ansatz,
predictions are obtained for the two scenarios of ordered resp.
unordered cascades. They have been derived from different
Monte Carlo models which either obey \kt ordering or do not underlie
such a restriction. The unordered scenario gives rise to a harder
\pt spectrum in the central rapidity region of the 
hadronic CMS than the ordered one.
It is further predicted that the \pt spectrum becomes harder resp.
softer with decreasing \xb for the unordered resp. ordered scenario. 
The application of the presented method at HERA
would not only allow to discriminate between
the different QCD models,
it would also offer the
possibility to resolve the question of \kt ordered vs. unordered cascade,
or DGLAP- vs. BFKL- like evolution at small~\xb.

\bigskip\bigskip
\noindent
{\bf Acknowledgements.}
{\footnotesize 
This work has been made possible by a grant from the Deutsche 
Forschungsgemeinschaft.
I would like to thank 
F. Botterweck and E. De Wolf for inspiring discussions,
and T. Carli and G. Grindhammer for their help with 
the Monte Carlo generators. 
A. De Roeck, R. Eichler, 
G. Grindhammer and G. Ingelman are gratefully 
acknowledged for their critical reading of the manuscript.}

\renewenvironment{thebibliography}[1]
	{\begin{list}{\arabic{enumi}.}
	{\usecounter{enumi}\setlength{\parsep}{0pt}
	 \setlength{\itemsep}{0pt} 
         \settowidth
	{\labelwidth}{#1.}\sloppy}}{\end{list}}

\end{document}